\documentclass[floatfix, noshowpacs, twocolumn, nofootinbib, aps, pra, longbibliography]{revtex4-1}

\usepackage{amsmath,amsthm,amssymb,amsfonts}
\usepackage{float}
\usepackage{graphicx}
\usepackage[colorlinks]{hyperref}
\usepackage{url}
\usepackage{mathtools}
\mathtoolsset{centercolon}
\usepackage{subdepth}
\usepackage{MnSymbol}
\usepackage{enumitem}

\newcommand{\bra}[1]{\left\langle{#1}\right\vert}
\newcommand{\ket}[1]{\left\vert{#1}\right\rangle}
\newcommand{\ketbra}[2]{\left\vert{#1}\right\rangle\left\langle{#2}\right\vert}
\DeclareMathOperator{\tr}{tr}

\newcommand{\swap}{\textsc{swap}}
\newcommand{\fswap}{\textrm{f-}\swap}
\newcommand{\creat}[1]{a_{#1}^{\dagger}}
\newcommand{\annih}[1]{a_{#1}^{~}}

\newcommand{\eq}[1]{Eq.~\hyperref[eq:#1]{(\ref*{eq:#1})}} 
\newcommand{\eqbeg}[1]{Equation~\hyperref[eq:#1]{(\ref*{eq:#1})}} 
\newcommand{\noteq}[1]{\hyperref[eq:#1]{(\ref*{eq:#1})}} 
\newcommand{\subeq}[1]{Eqs.~\hyperref[eq:#1]{(\ref*{eq:#1})}} 
\newcommand{\eqs}[2]{Eqs.~\hyperref[eq:#1]{(\ref*{eq:#1})} and \hyperref[eq:#2]{(\ref*{eq:#2})}} 
\newcommand{\eqsbeg}[2]{Equations \hyperref[eq:#1]{(\ref*{eq:#1})} and \hyperref[eq:#2]{(\ref*{eq:#2})}} 
\renewcommand{\sec}[1]{\hyperref[sec:#1]{Section~\ref*{sec:#1}}} 
\newcommand{\fig}[1]{\hyperref[fig:#1]{Fig.~\ref*{fig:#1}}} 
\newcommand{\thm}[1]{\hyperref[thm:#1]{Theorem~\ref*{thm:#1}}} 
\newcommand{\deftn}[1]{\hyperref[def:#1]{Definition~\ref*{def:#1}}} 
\newcommand{\deftns}[2]{Definitions \hyperref[def:#1]{\ref*{def:#1}} and \hyperref[def:#2]{\ref*{def:#2}}} 
\newcommand{\app}[0]{\hyperref[sec:appendix]{Appendix}} 

\theoremstyle{definition}
\newtheorem{theorem}{Theorem}
\newtheorem{definition}{Definition}

\begin{document}
\title{Efficient classical simulation of matchgate circuits with generalized inputs and measurements}

\author{Daniel J.\ Brod}
\email{dbrod@perimeterinstitute.ca}
\affiliation{Perimeter Institute for Theoretical Physics, 31 Caroline Street North, Waterloo, ON, N2L 2Y5, Canada}

\date{\today}
\begin{abstract}
Matchgates are a restricted set of two-qubit gates known to be classically simulable under particular conditions. Specifically, if a circuit consists only of nearest-neighbour matchgates, an efficient classical simulation is possible if either (i) the input is a computational basis state and the simulation requires computing probabilities of multi-qubit outcomes (including also adaptive measurements), or (ii) if the input is an arbitrary product state, but the output of the circuit consists of a single qubit. In this paper we extend these results to show that matchgates are classically simulable even in the most general combination of these settings, namely, if the inputs are arbitrary product states, if the measurements are over arbitrarily many output qubits, and if adaptive measurements are allowed. This remains true even for arbitrary single-qubit measurements, albeit only in a weaker notion of classical simulation. These results make for an interesting contrast with other restricted models of computation, such as Clifford circuits or (bosonic) linear optics, where the complexity of simulation varies greatly under similar modifications.
\end{abstract}
\maketitle

\section{Introduction} \label{sec:intro}

Matchgates are a class of restricted two-qubit gates with intriguing computational 
capabilities. Circuits composed of matchgates acting on 
nearest-neighbouring qubits (on a linear array) were shown to be classically 
simulable by Valiant \cite{Valiant2002}, and soon after shown to correspond to 
free fermions by Terhal and DiVincenzo \cite{Terhal2002}. Several other papers 
investigated the classical simulation of matchgates through different formalisms 
\cite{Knill2001a, Bravyi2005b, DiVincenzo2005, Jozsa2008b,Jozsa2008a}. However, 
matchgates also can become universal for quantum computation by the addition of 
seemingly simple resources. They were shown to be universal when supplemented by 
the $\swap$ gate \cite{Kempe2001b,Jozsa2008b}, by some two-qubit nondemolition measurements 
\cite{Beenakker2004}, by specific multi-qubit magic states \cite{Bravyi2006}, by 
almost any parity-preserving two-qubit gate \cite{Brod2011}, and on any 
connectivity graph that is not a path or a cycle \cite{Brod2012,Brod2014}.

In this paper, we are interested in how the complexity of simulating 
matchgates depends on restrictions on the inputs and outputs of the circuit. 
More concretely, we restrict our attention to circuits composed only of nearest-neighbour matchgates, 
and modify the computational model by allowing different types 
of input states and different restrictions on the size of the output. 
This is motivated by apparent differences between two previous results: that of 
Valiant \cite{Valiant2002}, and Terhal and 
DiVincenzo \cite{Terhal2002}, where the matchgate circuits act only on 
computational-basis inputs but any number of qubits can be measured at 
the end, and that of Jozsa and Miyake \cite{Jozsa2008b}, where the circuit can
act on arbitrary product inputs but the output consists of the measurement of a 
single qubit. Each of these settings was chosen with a specific application in mind,
 and it is not \emph{a priori} clear whether there 
is a common cause for the simulability of the different resulting computational 
models. 
Here we argue that it is indeed possible to unify these results---we show that matchgates can be simulated 
classically even if the input is in an arbitrary product state and the output 
consists of measurements of arbitrary subsets of the qubits, and this remains true even if one
is allowed to adapt subsequent gates depending on intermediate measurement outcomes.
By considering a weaker, sampling-based, classical simulation, we are also able to
extend these results to the case where measurements can be performed in arbitrary
single-qubit bases.

Besides refining our understanding of the computational power of matchgates, 
our results have other consequences that may be of more general interest. 
The first is that they provide a no-go result for some types of magic state injection 
protocols, namely if the magic states are single-qubit states. More 
specifically, universal quantum computation with nearest-neighbour 
matchgates is possible when certain auxiliary multi-qubit states are 
available \cite{Bravyi2006}. However, since we show that matchgates are simulable 
for arbitrary product inputs and adaptive measurements, this rules out a scheme 
similar to that of \cite{Bravyi2006} that only uses single-qubit ancillas. 
Noticeably, since the previous simulations  were restricted to either
computational basis inputs or single-qubit outputs, they could not be used to 
make this argument.

Our results can also be used to sharpen comparisons between matchgates and 
other restricted models of quantum computation. We will be especially 
interested in two examples: Clifford circuits and (bosonic) linear optics.

Clifford circuits are a particular class of quantum circuits widely 
known to be classically simulable under certain conditions \cite{Gottesman1999a}, 
with some similarities to matchgates 
\cite{Jozsa2008a}. However, several results have made it clear that the 
complexity of Clifford circuits is heavily dependent on the combined 
choices of inputs and outputs that the circuit has access to. The ``complexity 
landscape'' of Clifford circuits has recently been mapped out in 
\cite{Jozsa2014}, where the authors consider all combinations of: 
(i) computational basis 
versus arbitrary product inputs; (ii) single-qubit versus multi-qubit 
measurements; (iii) adaptive versus nonadaptive measurements; and (iv) weak 
versus strong simulation. The authors find that, by varying these conditions, 
the complexity of simulating Clifford circuits can go from (sub-)classical, to 
BQP-hard, to $\#$P-hard (cf.\ Figure 1 in \cite{Jozsa2014}). This has also been
extended to include arbitrary single-qubit measurements and different
notions of strong simulation \cite{Koh2015}. Our results consist, 
in a fashion, of a similar mapping of the complexity landscape of 
matchgate circuits, but with strikingly less diverse
results---matchgates are classically simulable in all possible combinations 
of the choices of \cite{Jozsa2014}, and almost all of those in \cite{Koh2015}.

Matchgates are also often compared to linear optics, due to a common underlying 
physical connection. While linear optics is identified with noninteracting 
bosons, matchgates are often identified with noninteracting fermions (and indeed, 
sometimes referred to as ``fermionic linear optics''). The mathematics behind linear 
optical circuits and matchgate circuits are surprisingly similar in some aspects 
(a point we will return to often throughout the paper, see also discussions in 
\cite{Terhal2002,Knill2001a}) but, while matchgate circuits are classically 
simulable, linear optics is not (see e.g.\ \cite{Knill2001b} for the KLM scheme 
for universal quantum computing with adaptive linear optics, or \cite{Aaronson2013a} for 
a model based on nonadaptive linear optics known as BosonSampling). 
However, these statements can be 
misleading if made without care---the separation in computational power between 
bosons and fermions is clear in the multi-qubit output setting, but if one is 
restricted to a single output measurement then bosonic linear optics can be simulated
\cite{Aaronson2014} in almost the same way as matchgates 
\cite{Jozsa2008a}. With the investigation we undertake here, we aim to shed 
further light on this comparison.

Finally, we believe that our results could also be used to inform 
the search for classical models of matchgates. More specifically, recent results 
have shown that both Clifford circuits \cite{Veitch2012} and linear-optical 
systems \cite{Veitch2013}, if constrained enough, can admit a classical probabilistic 
description. In other words, it is possible to construct hidden variable models for 
these systems which would preclude not only a computational speedup, but also other 
signature quantum features such as contextuality \cite{Spekkens2008}. The classical 
simulability of matchgates raises the natural question of whether a similar 
classical model can be constructed for these circuits, and the results 
obtained here could guide this search by suggesting sets of states
and measurements that are more likely to introduce nonclassical behaviours.

This paper is organized as follows. In \sec{backg} we give some preliminary 
definitions and background discussions. More specifically, in \sec{backgprel} we 
describe the Jordan-Wigner transformation and the mapping between matchgates and 
fermions, and in \sec{backsimul} we define a few different notions of classical 
simulation that we will need. In \sec{backgCIMO} we give a brief outline of the 
simulation obtained by Valiant \cite{Valiant2002} and Terhal and DiVincenzo 
\cite{Terhal2002}, and in \sec{backgPISO} we do the same for the simulation of 
Jozsa and Miyake \cite{Jozsa2008b}. In \sec{main} we prove our main result, which 
generalizes the two results discussed in the preceding sections, and discuss 
some possible extensions. We finish with some concluding remarks in 
\sec{summary}, as well as several open questions. The paper also contains an 
\app~with some further technical details omitted from the main text.

\textbf{Notation:} We will denote $X_i$, $Y_i$ and $Z_i$ the usual Pauli 
matrices acting on qubit $i$, and we will omit tensor product signs throughout. 
We will denote the anticommutator by $\{ A,B\} = AB + BA$. We will denote the 
all-zeroes state on $n$ qubits by $\ket{\bar{0}_n} = \ket{00\ldots0}$.

Throughout this paper, we will interchangeably refer to (unitary) quantum gates and 
their generating Hamiltonian. Since we will always be considering quantum computations in the 
circuit model (i.e.\ in a discrete-time description), whenever we refer to a 
gate by its generating Hamiltonian we in fact mean any unitary in the family generated 
by that Hamiltonian. 

Finally, throughout the paper we will use the following three acronyms to describe 
three types of circuit: CI--MO (computational input and multi-qubit output), 
PI--SO (product input and single-qubit output) and PI--MO (product input and 
multi-qubit output). The precise corresponding definitions can be 
found in \sec{backsimul}.

\section{Background} \label{sec:backg}

\subsection{Preliminary definitions: the Jordan-Wigner transformation} \label{sec:backgprel}

Let us begin with the following definition. 

\begin{definition} \label{def:matchgates}
    (Matchgates) Let $G(A,B)$ be the two-qubit gate given by
    \begin{equation} \label{eq:Matchgate}
        G(A,B) = 
        \begin{pmatrix}
            A_{11} & 0 & 0 & A_{12} \\
            0 & B_{11} & B_{12} & 0 \\
            0 & B_{21} & B_{22} & 0 \\
            A_{21} & 0 & 0 & A_{22}
        \end{pmatrix}.
    \end{equation}
    Then $G(A,B)$ is a \emph{matchgate} if $\det A = \det B$.
\end{definition}

The set of all two-qubit gates $G(A,B)$ acting on qubits $\{i,j\}$ corresponds 
to those generated by

\begin{equation} \label{eq:matchgroup}
    \mathcal{A}_{i,j} = \{ X_i X_j, X_i Y_j, Y_i X_j, Y_i Y_j, Z_i, Z_j \}
\end{equation}
It is well-known that the operators in $\mathcal{A}_{i,i+1}$ are closely connected 
to the physics of noninteracting fermions. To see that, let us define the 
following Jordan-Wigner operators \cite{Jordan1928} acting on $n$ qubits:
\begin{subequations} \label{eq:JWas}
    \begin{align} 
        \creat{j} &:= \left( \prod_{k=1}^{j-1} Z_k \right) \left(\frac{X_j-i Y_j}{2}\right), \\
        \annih{j} &:= \left( \prod_{k=1}^{j-1} Z_k \right) \left(\frac{X_j+i Y_j}{2}\right),
    \end{align}
\end{subequations}
for $j \in \{1,\ldots,n\}$. These operators satisfy the anti-commutation 
relations one would expect for fermionic operators:
\begin{subequations} \label{eq:anticomm}
    \begin{align}
        \{\creat{i},\creat{j}\} & = 0, \\
        \{\annih{i},\annih{j}\} & = 0, \\
        \{\annih{i},\creat{j}\} & = \delta_{i,j},
    \end{align}
\end{subequations} 
for all $i, j \in \{1,\ldots,n\}$. If we identify 
states $\ket{0}$ and $\ket{1}$ of qubit $i$ with the empty and occupied 
states of fermionic mode $i$, respectively, then $\creat{i}$ ($\annih{i}$) 
behaves precisely as a fermionic creation (annihilation) operator. From 
\subeq{JWas} we also obtain
\begin{align}
    Z_k =(\creat{k}-\annih{k}) (\annih{k}+\creat{k}), \label{eq:JWtransf1}
\end{align}
for $k \in \{1,\ldots,n\}$, and
\begin{subequations} \label{eq:JWtransf2}
    \begin{align}
        X_k X_{k+1} & = - (\annih{k}-\creat{k})(\annih{k+1}+\creat{k+1}), \\
        Y_k Y_{k+1} & = (\annih{k}+\creat{k})(\annih{k+1}-\creat{k+1}),  \\
        Y_k X_{k+1} & = i (\annih{k}+\creat{k})(\annih{k+1}+\creat{k+1}), \\
        X_k Y_{k+1} & = i (\annih{k}-\creat{k})(\annih{k+1}-\creat{k+1}), 
    \end{align}
\end{subequations}
for $k \in \{1,\ldots,n-1\}$. \eqsbeg{JWtransf1}{JWtransf2} connect  the 
generators of {\em nearest-neighbour} 
matchgates, $\mathcal{A}_{i,i+1}$, precisely to quadratic fermionic Hamiltonians.

To avoid ambiguity, we should point out that the notion of locality
is not preserved by the Jordan-Wigner transformation. In 
particular, a quadratic operator acting between distant fermionic modes, e.g.\ 
$(\annih{1}-\creat{1})(\annih{3}+\creat{3})$, maps to the multi-qubit operator 
$X_1 Z_2 X_3$, not to the two-qubit matchgate $X_1 X_3$. 
In fact, the most general multi-qubit operators obtained from quadratic 
fermionic operators are $A_i Z_{i+1} Z_{i+2} \ldots Z_{j-1} B_j$, for $i < j$, 
where $A$ and $B$ are either $X$ or $Y$. Since any such Hamiltonian can 
be implemented by a poly-sized circuit of nearest-neighbour 
matchgates\footnote{This is the fermionic analogue of the well-known fact that 
any photonic interferometer can be decomposed in terms of O$(n^2)$ 
nearest-neighbour beam splitters \cite{Reck1994}.}, as shown in 
\cite{Jozsa2008b}, they are (computationally) equivalent to nearest-neighbour 
matchgates. In contrast, almost any gate generated by $\mathcal{A}_{i,j}$ where 
$i$ and $j$ are non-neighbouring qubits leads to universal quantum computation 
\cite{Brod2014}. In light of these considerations, throughout this paper we 
will, unless stated otherwise, restrict our attention to circuits of 
\emph{nearest-neighbour} matchgates, in the qubit picture, or quadratic 
fermionic operators between \emph{arbitrary pairs of modes}, in the fermionic 
picture, keeping in mind that these are computationally equivalent. 

A consequence of this observation, which will be useful later
on, is that the overall ordering of the qubits is irrelevant. More specifically,
given any circuit of nearest-neighbour matchgates $M$, we can find the corresponding
transformation in the fermionic picture, apply some permutation $P$ on the labels
of the fermionic modes, then map everything back to the qubit picture to obtain 
a different circuit $M'$. But, by the considerations of the previous paragraph,
the new circuit $M'$ can be decomposed as a circuit of matchgates with only polynomial
overhead, and furthermore these matchgates now act between nearest-neighbours according
to the relabelling of the qubits induced by the permutation $P$.

An important property of quadratic gates, which is crucial to the 
classical simulation schemes that follow, is that they act \textit{linearly} on 
creation and annihilation operators (hence matchgates are often called 
fermionic linear optics). More specifically, if $M$ is an unitary operator 
corresponding to a 
circuit of nearest-neighbour matchgates, then we can write (for a simple proof, 
see \cite{Jozsa2008b}):
\begin{equation} \label{eq:linearoptics}
    M \creat{i} M^{\dagger} = \sum_{j=1}^n R_{i j} \creat{j} + \sum_{j=1}^n {R'}_{i j}\annih{j}.
\end{equation}
If $M$ further only consists of ``number-preserving'' matchgates, i.e.\ those 
$G(A,B)$ for which $A$ is diagonal\footnote{Alternatively, quadratic operators 
restricted to the combinations $\annih{j} \creat{k}$ and $\annih{k} \creat{j}$, 
or matchgates generated by $X_k X_{k+1} + Y_k Y_{k+1}$, $X_k Y_{k+1} - Y_k 
X_{k+1}$, and $Z_k$.}, then ${R'}=0$. Curiously, an analogous version of 
\eq{linearoptics} also holds for \textit{bosonic} linear optics---thus we expect 
that, even if \eq{linearoptics} is behind the classical simulability of 
matchgates, it cannot be the whole story. We will return to this point several 
times as we discuss the different types of simulation results throughout this 
Section.

\subsection{Preliminary definitions: classical simulation} \label{sec:backsimul}

Before moving to our main result, let us define precisely
what is meant by classical simulation. In particular, suppose our model of 
computation consists of an uniform family of quantum circuits, $\{C_n\}$, 
which act on yet-unspecified $n$-qubit input states $\ket{\psi_n}$. Suppose also that 
the circuits are followed by measurements of some subset of $k$ out of the $n$ qubits 
in the computational basis. Then, for any $k$-bit string $\tilde{y}$ corresponding to some assignment of the $k$ measured qubits, we write the 
probability of observing measurement outcome $\ket{\tilde{y}}$ as
\begin{equation} \label{eq:probdef}
    \textrm{Pr}(\tilde{y}|\psi_n) = \tr \bra{\tilde{y}}C_n\ketbra{\psi_n}{\psi_n}C_n^{\dagger} \ket{\tilde{y}},
\end{equation}
where the partial trace is taken over the unmeasured qubits. We can now divide our 
notions of classical simulation in a few convenient types (this is not an 
exhaustive list, see \cite{Nest2010, Bremner2011} for more detailed 
discussions):

\begin{definition} \label{def:strongsimul}
    (Strong simulation) The uniform family of quantum circuits $\{C_n\}$, acting
     on the $n$-qubit input state $\ket{\psi_n}$, is \emph{strongly simulable} if, 
     for every assignment of $k$ output qubits $\tilde{y}$, and for every $k$, 
     it is possible to compute Pr$(\tilde{y}|\psi_n)$ to $m$ digits of precision in 
     poly$(n,m)$ time on a classical computer.
\end{definition}

\begin{definition} \label{def:weaksimul}
    (Weak simulation) The uniform family of quantum circuits $\{C_n\}$, acting on the 
    $n$-qubit input state $\ket{\psi_n}$, is \emph{weakly simulable} if, for every choice
    of $k$ out of $n$ qubits to be measured, for every $k$, it is possible to produce a sample 
    from the probability distribution defined by Pr$(\tilde{y}|\psi_n)$ in poly$(n)$ 
    time on a classical computer.
\end{definition}

Note that, as defined, strong simulation implies weak simulation\footnote{But 
not the other way around, as there are examples for which weak simulation is 
easy, but strong simulation is $\#$P-hard \cite{Nest2010}.}. Weak simulation is 
often considered more physically-motivated, since any quantum device only 
outputs samples from a probability distribution, and requiring a classical 
device to compute the probabilities to high precision does not make for a 
fair comparison of their respective computational powers. On the other hand, 
for part of the cases considered in this paper it will be simple enough to prove that 
strong simulation is possible. We will also define two variants of the above:

\begin{definition} \label{def:singlequbitsimul}
    (Single-output strong simulation) The uniform family of quantum circuits 
    $\{C_n\}$, acting on the $n$-qubit input state $\ket{\psi_n}$, is 
    \emph{strongly simulable with a single output} if the quantity 
    $\bra{\psi_n} C_n^{\dagger} Z_i C_n \ket{\psi_n}$, for any $1 \leq i \leq n$,
     can be computed to $m$ digits of precision in poly$(n,m)$ time on a classical 
     computer. Note that $\bra{\psi_n} Z_i \ket{\psi_n} =p_i(0) - p_i(1)$, 
     where $p_i(j)$ is the probability that qubit $i$ will be measured in 
     state $\ket{j}$.
\end{definition}

This definition is useful if one wants to characterize some restricted 
computational model in terms of the \emph{decision problems} it can solve 
(i.e., problems with a single YES or NO answer), where the answer to the 
problem is encoded in a single output qubit. 

\begin{definition} \label{def:adaptivesimul}
    (Adaptive simulation) Let $\{C_n\}$ be a uniform family of \emph{adaptive} 
    quantum circuits, that is, quantum circuits where one is allowed to make 
    intermediate measurements and condition subsequent operations on their 
    outcomes. Then  $\{C_n\}$, acting on the $n$-qubit input state $\ket{\psi_n}$, 
    is \emph{adaptively simulable} if (i) all intermediate measurements can be 
    weakly simulated (in the sense of \deftn{weaksimul}), and (ii) the final
    measurements on the circuit determined by the outcomes of (i) can be 
    strongly simulated.
\end{definition}

We presented this hybrid definition of classical simulation to capture more 
closely the workings of an adaptive protocol: the complete circuit is not known 
at the beginning of the computation, as it depends on intermediate measurement 
outcomes. Then \deftn{adaptivesimul} requires the classical computer to 
randomly choose the outcomes of intermediate measurements according to the correct 
distribution and, after the complete circuit is determined, to calculate 
the probabilities of the computational outcomes\footnote{Note that \deftn{adaptivesimul} does
\emph{not} require the classical computer to strongly simulate the final measurement outcomes
of $C_n$, which would correspond to computing the average probabilities of the final measurements
weighed by the probabilities of all possible intermediate outcomes.}.
This would be unnecessary if we had a universal set of 
quantum gates at hand---we could simply replace measurement adaptations by 
coherently controlled gates, defer all intermediate measurements to the end of 
the circuit, and perform a strong simulation of the resulting circuit 
\cite{LivroNielsen}. However, these controlled gates might not be available in a 
given restricted model, and in fact measurement adaption plays an important role 
in several models of quantum computation, most notably bosonic linear optics 
\cite{Knill2001b}, Clifford circuits with magic state injection 
\cite{Bravyi2005a}, and measurement-based quantum computation 
\cite{Raussendorf2001}.

Since the main focus of this work is the interplay between restrictions in the 
inputs and measurements of the circuits, we also define the following 
nomenclature. A computational input / multi-qubit output, or CI--MO, simulation 
is a restriction of Definitions 
\hyperref[def:strongsimul]{\ref*{def:strongsimul}}, 
\hyperref[def:weaksimul]{\ref*{def:weaksimul}} or 
\hyperref[def:adaptivesimul]{\ref*{def:adaptivesimul}} to the case where the 
input state, $\ket{\psi_n}$, is just a computational basis state $\ket{x}$ for 
some bit string $x$. A product input / single-qubit output, or PI--SO, 
simulation is a restriction of \deftn{singlequbitsimul} to the case where the 
input $\ket{\psi_n}$ is an arbitrary product state. Finally, a product input / 
multi-qubit output simulation, or PI--MO, is the natural extension where the 
input can be an arbitrary product state and the measurements are over any subset 
of the qubits.

\subsection{CI--MO simulation of matchgates} \label{sec:backgCIMO}

Let us now describe the CI--MO simulation of matchgates due to Valiant 
\cite{Valiant2002}, and Terhal and DiVincenzo \cite{Terhal2002} (for convenience 
we will follow more closely the latter). We begin by stating:

\begin{theorem}\label{thm:CIMO}
    (\cite{Valiant2002,Terhal2002}). Let $\{M_n\}$ be a uniform family of (possibly 
    adaptive) quantum circuits composed of poly$(n)$ nearest-neighbour matchgates 
    acting on $n$ qubits, and let the input to the circuit be a state $\ket{x}$ for 
    any $n$-bit string $x$. Then, there are polynomial-time classical algorithms to 
    simulate the outcomes of measurements over arbitrary subsets
    of the output qubits in the weak, strong and adaptive sense.
\end{theorem}

In the \app~we outline the proof of \thm{CIMO} for the particular case of 
``number-preserving'' matchgates [i.e., when ${R'}=0$ in \eq{linearoptics}]. The 
crucial property of matchgates that makes \thm{CIMO} true is the fact that all 
outcome probabilities [cf.\ \eq{probdef}] can be written in terms of 
\emph{matrix determinants}. For example, if $y$ and $\tilde{y}$ are arbitrary 
$n$-bit and $k$-bit strings, respectively, corresponding to a total or partial assignment of the output qubits, we can write
\begin{align}
    \textrm{Pr}(y|x) = & |\bra{y} M_n \ket{x}|^2 = |\det (R_{x,y})|^2, \label{eq:fermionsamplingCIMO}\\
    \textrm{Pr}(\tilde{y}|x) = & \tr \bra{\tilde{y}}M_n\ketbra{x}{x}M_n^{\dagger}\ket{\tilde{y}} = \textrm{Pf} (\tilde{M}). \label{eq:probystarCIMO}
\end{align}
where $R_{x,y}$ is a specific submatrix of the matrix $R$ from 
\eq{linearoptics}, and $\tilde{M}$ is a poly-sized antisymmetric matrix 
constructed out of the matrix elements of $R$ in a specific manner. We direct 
the interested reader to the \app~for a description of the intuition behind 
these expressions, or to the original paper \cite{Terhal2002} for 
lookup tables that explain how to construct $R_{x,y}$ and $\tilde{M}$. The Pfaffian Pf$(A)$, 
that appears in \eq{probystarCIMO}, is a matrix polynomial that, for an $n 
\times n$ antisymmetric matrix $A$, is 0 if $n$ is odd and satisfies the 
relation
\begin{equation*}
    \textrm{Pf} (A)^2 = \det(A)
\end{equation*}
if $n$ is even. In the \app~we give a small generalization of \thm{CIMO}, 
showing that it holds also for periodic boundary conditions (i.e.\ if matchgates can 
also act between the first and last qubits).

As discussed in \sec{backgprel}, although the linearity of \eq{linearoptics} 
seems important for the simulation of matchgates, it is indeed not the 
whole story: for a CI--MO simulation, the probabilities in 
\eqs{fermionsamplingCIMO}{probystarCIMO} involve, a priori, the sum of an 
exponentially-large number of terms, however the final expressions coalesce 
into easy-to-compute determinants. In fact, it is interesting to contrast this 
CI--MO simulation of matchgates to their bosonic counterpart. 
Bosonic linear optics includes BosonSampling \cite{Aaronson2013a}, a model for which there
is strong evidence that an efficient classical simulation is impossible,
and when imbued with adaptive measurements it is capable of universal 
quantum computation \cite{Knill2001b}. Thus bosons apparently display 
a great computational advantage over fermions, and this seems consequence of the fact
that, rather than determinants (or Pfaffians), bosonic evolution is described by 
permanents, which are dramatically harder to compute (in fact, among the hardest 
problems in the complexity class $\#$P \cite{Valiant1979}).

\subsection{PI--SO simulation of matchgates} \label{sec:backgPISO}

For completeness, we now provide a brief outline of the simulation scheme used 
e.g.\ by Jozsa and Miyake in \cite{Jozsa2008b}, although our main result in 
following sections will be based on \thm{CIMO}. We begin by stating:

\begin{theorem}\label{thm:PISO}(\cite{Jozsa2008b}).
    Let $\{M_n\}$ be a uniform family of quantum circuits composed of poly$(n)$ 
    nearest-neighbour matchgates acting on $n$ qubits, and let the input be an 
    arbitrary $n$-qubit product state 
    $\ket{\psi} = \ket{\psi_1} \ket{\psi_2} \ldots \ket{\psi_n}$. 
    Then we can efficiently compute the expectation value 
    $\langle Z_k \rangle = \bra{\psi} M_n^{\dagger} Z_k M_n \ket{\psi}$,
    i.e., there is an efficient strong simulation in the single-output sense 
    of \deftn{singlequbitsimul}.
\end{theorem}

\thm{PISO} is a consequence of the linearity of 
\eq{linearoptics}. First note that, by \eq{JWtransf1}, we can write $Z_k = 
\creat{k} \annih{k} - \annih{k}\creat{k}$. But then, by \eq{linearoptics} there 
are $R$ and $R'$ such that
\begin{align} \label{eq:PISO}
    & \bra{\psi} M_n^{\dagger} \creat{k} \annih{k} M_n \ket{\psi}=\notag \\
    & \bra{\psi} \sum_{j,l}^n \left( R_{k j} \creat{j} + {R'}_{k j} \annih{j}     
    \right) \left( R^{\textrm{*}}_{k l} \annih{l} + {R'}^{\textrm{*}}_{k l}\creat{l} \right)\ket{\psi}
\end{align}
and similarly for $\bra{\psi} M_n^{\dagger} \annih{k} \creat{k} M_n 
\ket{\psi}$. \eqbeg{PISO} consists of a sum of a polynomial number of terms 
of the type $\bra{\psi} \creat{j} \annih{k} \ket{\psi}$ for all 
quadratic combinations of creation and annihilation operators. But, from 
\eqs{JWtransf1}{JWtransf2} and subsequent discussion, all such quadratic terms 
are tensor products of Pauli matrices. Since 
$\ket{\psi}$ is a product state, all expectation values that appear in 
\eq{PISO} factor into products of single-qubit expectation values of Pauli 
matrices. Thus, it is clear $\langle Z_k \rangle$ can be computed with only 
poly($n$) computational effort, which essentially proves \thm{PISO}. This result 
was further extended in \cite{Jozsa2008b} to allow for measurement of a 
logarithmic-sized subset of the output qubits, and in \cite{Brod2014} to allow 
for periodic boundary conditions.

In contrast to \thm{CIMO}, the proof of \thm{PISO} seems to rely on the 
fact that \eq{linearoptics} is a linear transformation between creation and 
annihilation operators rather than on any intrinsically fermionic property. 
This is further supported by the fact that, for bosonic linear 
optics, a similar quantity to $\langle Z_k \rangle$ can also be computed 
efficiently \cite{Knill2003}, and it is easy to sample classically from a 
BosonSampling distribution if we're restricted to a single output mode 
\cite{Aaronson2014}. The proof of both facts also seem to stem from the 
linearity of \eq{linearoptics}. 

The interpretation of the single-output setting of \thm{PISO} in terms of 
decision problems has also led to interesting mappings between matchgate 
circuits and (classical) circuits of linear threshold gates \cite{Nest2011a}, or 
between matchgate circuits and arbitrary logspace quantum computers [i.e.\ 
universal circuits acting on O$($log$\,n$) qubits] \cite{Jozsa2010}. The latter result also led to 
novel proposals for compressed simulation of spin systems on small-scale quantum 
computers \cite{Kraus2011, Boyajian2013}. Since there seems to be less 
difference between bosons and fermions in the single-output setting, an 
interesting question arises of whether the results of 
\cite{Jozsa2010,Nest2011a,Kraus2011,Boyajian2013} could have some nontrivial bosonic analogue.

\section{Main result: efficient PI--MO simulation of matchgate circuits} \label{sec:main}

The comparisons between fermionic and bosonic linear optics at the ends of 
\sec{backgCIMO} and \sec{backgPISO} seem to suggest that efficient PI--SO and 
CI--MO simulations of matchgates are possible for fundamentally different 
reasons---the former is a consequence of fermionic probabilities being described 
by determinants, whereas the latter seems to be a consequence of the linear 
relation satisfied by free particles [i.e.\ \eq{linearoptics}], and in fact only 
the latter seems possible for free bosons.

In this section, we argue that this apparent difference is not fundamental. 
More specifically, we show how to extend the result of 
\cite{Terhal2002} to allow efficient classical simulation of matchgate circuits 
with arbitrary product inputs and measurements of arbitrary subsets of the 
output in the computational basis (that is, a PI--MO simulation). 

We begin by stating the following theorem:

\begin{theorem}\label{thm:PIMO}
    Let $\{M_n\}$ be a uniform family of (possibly adaptive) quantum circuits 
    composed of poly$(n)$ nearest-neighbour matchgates acting on $n$ 
    qubits, and let the input be an arbitrary $n$-qubit product state $
    \ket{\psi} = \ket{\psi_1} \ket{\psi_2} \ldots \ket{\psi_n}$. 
    Then, there are polynomial-time classical algorithms to simulate the 
    corresponding outcomes in the weak, strong and adaptive sense.
\end{theorem}

The first step to prove \thm{PIMO} is to replace the arbitrary 
product state $\ket{\psi} = \ket{\psi_1} \ket{\psi_2} \ldots 
\ket{\psi_n}$ by a circuit of matchgates acting on a fiducial state. To that 
end, we will use the following identities (see e.g.\ 
\cite{Brod2012,Brod2014}):
\begin{subequations} \label{eq:ids}
    \begin{align}
        G(H,H) \ket{\phi} \ket{+} & = (H \ket{\phi}) \ket{+} \label{eq:id1}, \\
        G(Z,X) \ket{\phi} \ket{0} & = \ket{0} \ket{\phi} \label{eq:id2}, \\
        G(Z,X) \ket{0} \ket{\phi} & = \ket{\phi} \ket{0} \label{eq:id3},
    \end{align}
\end{subequations}
where $\ket{\phi}$ is an arbitrary single-qubit state and $H$ is the usual
single-qubit Hadamard matrix. \eqbeg{id1} means that 
$G(H,H)$ can induce an $H$ gate on a qubit state $\ket{\phi}$ when 
it has access to an ancilla in the $\ket{+}$ state, and \eqs{id2}{id3} mean that 
the fermionic $\swap$ gate, defined as $\fswap := G(Z,X)$, behaves exactly as 
the $\swap$ gate when one of the qubits is in the $\ket{0}$ state. These 
identities are useful because neither $H$ nor $\swap$ are matchgates on their 
own. In fact, either gate, when added to the set of matchgates, leads to 
universal quantum computation \cite{Terhal2002,Jozsa2008b}, and so we clearly do 
not expect to be able to replace them by matchgates in general. Nevertheless, 
\subeq{ids} show how to do this in some particular 
cases by a suitable use of ancilla states\footnote{These simple identities 
provide quite a lot of leverage, and were crucial to show that matchgates are 
universal when acting on almost all connectivity graphs in 
\cite{Brod2012,Brod2014}.}.

\begin{figure}[t]
    \centering
    \includegraphics[width=0.45\textwidth]{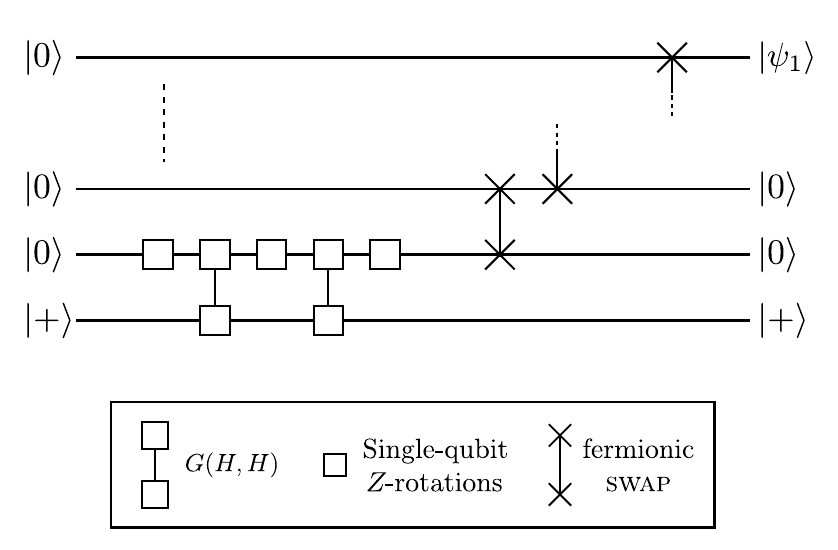}
    \caption{By adding a $\ket{+}$ ancilla at the end of the circuit to act as a 
    catalyst, it is possible to sequentially prepare the qubits in an arbitrary 
    product state.}
    \label{fig:replaceinput}
\end{figure} 

Consider now the circuit of \fig{replaceinput}. By repeated application of 
\subeq{ids}, it starts from the $(n+1)$-qubit state 
$\ket{\bar{0}_n}\ket{+}$ and prepares the desired state $\ket{\psi_1} 
\ket{\psi_2} \ldots \ket{\psi_n} \ket{+}$ via the following procedure:

\begin{itemize}[leftmargin=0.7 cm]
    \item[(i)] Use the $\ket{+}$ ancilla to apply $H$ gates to qubit $n$ via \eq{id1} 
    which, together with single-qubit $Z$ rotations (matchgates 
    themselves), can be used to prepare qubit $n$ in state $\ket{\psi_1}$;
    \item[(ii)] Since all qubits from $1$ to $n-1$ are initially in the $\ket{0}$ 
    state, use \eq{id3} to effectively $\fswap$ the state of 
    qubit $n$ all the way up to qubit $1$.
    \item[(iii)] At this point, we have the state $\ket{\psi_1} \ket{\bar{0}_{n-1}} 
    \ket{+}$;
    \item[(iv)] Repeat steps (i)-(iii) to sequentially prepare each state $
    \ket{\psi_i}$ and $\fswap$ it to the qubit at position $i$.
\end{itemize}

After following steps (i)-(iv) we are left with the state $\ket{\psi_1} 
\ket{\psi_2} \ldots \ket{\psi_n} \ket{+}$. From this point on we can ignore 
qubit $n+1$ and perform the original matchgate circuit $M_n$ from \thm{PIMO}. 

The procedure above allows us to replace the initialization of any input product 
state by the initialization of a standard input state, $\ket{\bar{0}_n}\ket{+}$, 
followed by the matchgate circuit of \fig{replaceinput}, which we denote by $U$. 
Our claim is that it is possible to compute, with only twice the computational 
effort, the same quantities as in the CI--MO simulation of \sec{backgCIMO}. We can do 
this by applying the same methods to the circuit $M_n U$, although this is not 
immediately apparent since the input in \fig{replaceinput} is not in the 
computational basis. To show how this can be circumvented, let $\tilde{y}$ be 
some assignment of a subset of $k$ out of the $n$ qubits, for any $k \leq n$, and write
\begin{align} \label{eq:probystarPIMO}
    \textrm{Pr}(\tilde{y}|\psi) & =  \bra{\psi} M_n^{\dagger} P_{\tilde{y}} M_n 
\ket{\psi} \notag \\
    & =\!\frac{1}{2} \bra{\bar{0}_{n+\!1}} (1\!+\!\annih{n+\!1}) U^{\dagger} 
M_n^{\dagger}  P_{\tilde{y}} M_n U (1\!+\!\creat{n+\!1}) 
\ket{\bar{0}_{n+\!1}}\!,
\end{align}
where we rewrote state $\ket{+}$ as fermionic 
operators acting on $\ket{\bar{0}_{n+1}}$. Here, $P_{\tilde{y}}:=\ketbra{\tilde{y}}{\tilde{y}}$ 
is a projector that can be written as a string of creation and annihilation operators as 
follows. First, label the $k$ qubits assigned by $\tilde{y}$ as 
$\{l_1, l_2, \ldots, l_k\}$. Then, for each bit $\tilde{y}_i$ assigned 
to qubit $l_i$, choose either $(\creat{l_i} \annih{l_i})$ if it is 1 
or $(\annih{l_i} \creat{l_i})$ if it is 0. Finally, define $P_{\tilde{y}}$ as the product of these operators. 
For example, one could obtain $P_{\tilde{y}}=(\creat{l_1} \annih{l_1})(\creat{l_2} \annih{l_2}) \ldots (\annih{l_k} \creat{l_k})$, for some bit string $\tilde{y} = 11\ldots0$.
 
Following the same steps that lead to \eq{probystarCIMO} (which we omit, but are outlined 
in the~\app~and worked out in full detail in \cite{Terhal2002}) we can obtain
\begin{align} \label{eq:probystarPIMO2}
    \textrm{Pr}(\tilde{y}|\psi) = & \frac{1}{2} \bra{\bar{0}_{n+1}} U^{\dagger} 
M_n^{\dagger}  P_{\tilde{y}} M_n U \ket{\bar{0}_{n+1}} \notag \\
   + & \frac{1}{2} \bra{\bar{0}_{n+1}} \annih{n+1} U^{\dagger} M_n^{\dagger}  
P_{\tilde{y}} M_n U \creat{n+1} \ket{\bar{0}_{n+1}} \notag \\
   = & \frac{1}{2} (\textrm{Pf} (\tilde{M}_1)+\textrm{Pf} (\tilde{M}_2)),
\end{align}
where $\tilde{M}_1$ and $\tilde{M}_2$ are defined as in \eq{probystarCIMO}, and 
can be easily constructed using the lookup tables found in \cite{Terhal2002}. 
Intuitively, this simplification is possible because matchgates preserve parity, 
and so the combined circuit $M_n U$ acts independently on 
$\ket{\bar{0}_{n}}\ket{0}$ and $\ket{\bar{0}_{n}}\ket{1}$, thus these two parity 
branches never interfere. 

From \eq{probystarPIMO2} and \fig{replaceinput}, it is clear that the strong 
simulation of \thm{PIMO} is possible, since tracking the parallel evolution of the 
two parity branches of the state reduces to simulating two independent CI--MO 
instances, as per \thm{CIMO}.

The circuit of \fig{replaceinput} is closely related to another trick, used in 
\cite{Knill2001a, Jozsa2015}, where one maps \emph{linear} fermionic operators 
[i.e.\ those in \subeq{JWas}] on $n$ fermionic modes to quadratic operators 
[i.e.\ those in \eqs{JWtransf1}{JWtransf2}] on $n+1$ fermionic modes, by adding 
one ancilla mode. Even so, the authors of 
\cite{Knill2001a, Jozsa2015} only considered either PI--SO or CI--MO settings.

Another surprising aspect of this construction is the fact that the two parity 
branches of state $\ket{\psi}$ can be obtained using nearest-neighbour 
matchgates from a superposition of the simplest bit strings of different 
parities. Since matchgates preserve parity, we should of course have expected 
that the two parity branches of $\ket{\psi}$ would evolve independently. 
Nonetheless, if we start from an arbitrary product state $\ket{\psi}$ and look 
at its projection onto the even parity subspace, say, we are left with a 
complicated entangled state, and it is not obvious that it would have an 
efficient description that would preserve the classical simulability of 
matchgates. The circuit of \fig{replaceinput} shows that this is in 
fact the case.

Finally, let us show why the adaptive simulation of \thm{PIMO} is possible, 
using a similar argument as for the CI--MO case [cf.\ the discussion surrounding 
\eq{probystarADAP}]. For simplicity, suppose the whole adaptive circuit $M_n$ we 
wish to simulate consists of (i) an $n$-qubit matchgate circuit $M$, (ii) 
measurement of a single qubit $y_1$, and (iii) either of two matchgate circuits, 
which we represent by $M_{y_1}$ depending on the outcome of $y_1$. As before, this 
circuit acts on some input product state $\ket{\psi} = \ket{\psi_1} \ket{\psi_2} 
\ldots \ket{\psi_n} \ket{+}$, and at the end we wish to compute the probability 
of the $k$-bit string $\tilde{y}_2$ on some assignment of $k$ out of the $n-1$ 
remaining qubits.

To do this, we first replace the state $\ket{\psi}$ by the circuit of 
\fig{replaceinput} acting on $\ket{\bar{0}_n}\ket{+}$, as before. Then we 
perform the simulation as follows:

\begin{itemize}[leftmargin=0.7 cm]
    \item[(i)] Compute $\textrm{Pr}(y_1|\psi)$ by applying \eq{probystarPIMO2};
    \item[(ii)] Classically sample according to the probabilities computed in (i), 
    and fix the corresponding outcome for $y_1$.
    \item[(iii)] Compute Pr$({\tilde{y}_2}, y_1 | \psi)$, given by
    \begin{equation} \label{eq:probystarPIMOADAP}
        \begin{aligned}
       & \frac{1}{2} \bra{\bar{0}_{n+\!1}} U^{\dagger} M^{\dagger} P_{y_1} M_{y_1}^{\dagger} P_{\tilde{y}_2} M_{y_1} P_{y_1} M U \ket{\bar{0}_{n+\!1}}   \\
         + & \frac{1}{2} \bra{\bar{0}_{n+\!1}} \annih{n+1} U^{\dagger} M^{\dagger} P_{y_1} M_{y_1}^{\dagger} P_{\tilde{y}_2}  M_{y_1} P_{y_1} M U \creat{n+1} \ket{\bar{0}_{n+\!1}},
        \end{aligned}
    \end{equation}
   where, again, both $P_{y_1}$ and $P_{\tilde{y}_2}$ are strings of creation and 
   annihilation operators determined by the assignments $y_1$ and 
   $\tilde{y}_2$, respectively, as done for \eq{probystarPIMO}.
   Clearly, the main difference between \eq{probystarPIMOADAP} and 
   \eq{probystarPIMO2} is the introduction of the projector $P_{y_1}$ between the two 
   parts of the circuit. But this operator is also even in the fermionic 
   operators, so the same argument as before applies, and the probability factors as 
   the sum of the probabilities of two independent (adaptive) CI--MO simulations. 
\end{itemize}

Another way to state this result is that a projective measurement of a single 
qubit on the computational basis is itself a parity-preserving operation, so the 
adaptive measurement preserves the structure of two parallel simulations of 
matchgate circuits acting on well-defined parity states. Clearly, one can extend 
this simulation to allow for a polynomial number of rounds of measurements on 
different subsets of qubits, such as done for the CI--MO case in 
\cite{Terhal2002}.

\subsection{Measurement on non-computational bases} \label{sec:noncomputational}

After extending the results of classical simulability of matchgates to 
include arbitrary input product states, the next natural question that arises is 
whether we can also change the \emph{measurements} to allow for arbitrary 
non-computational-basis measurements. Conceptually, this could be framed as 
an even stronger simulation than that of \deftn{strongsimul}, 
since we would be able to compute the probabilities of a 
\emph{tomographically-complete} set of measurements. (For comparison, 
note that this is possible for 
Clifford circuits in all cases where they are strongly simulable, since they include 
the gates that map the computational basis to the $X$ and $Y$ basis.)

Currently, it is not clear how to perform this simulation for the most 
general single-qubit measurements, or even only in a tomographically complete set of 
measurement bases. Short of that, we will show how to perform a \emph{weak} simulation 
of the circuits (cf.\ \deftn{weaksimul}). Although this provides a much less 
precise description of the output state, it already suffices to rule out the possibility 
that matchgates could leverage arbitrary single-qubit measurements to perform 
universal quantum computation.

\begin{figure}[t]
    \centering
    \includegraphics[width=0.45\textwidth]{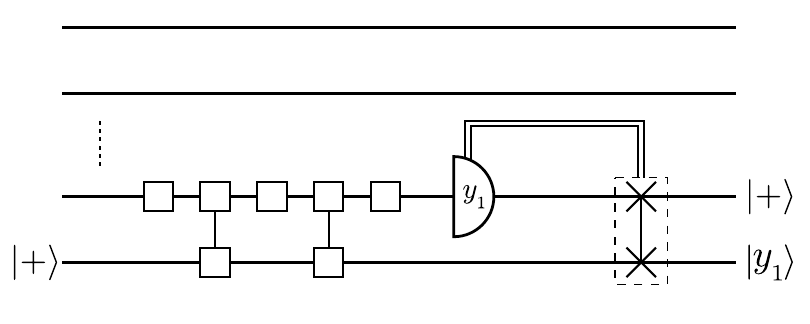}
    \caption{This circuit rotates the measurement basis of the last qubit using a 
$\ket{+}$ ancilla and a sequence of matchgates. The outcome controls classically 
control which variant of the $\fswap$ gate, $G(Z,X)$ or $G(-Z,X)$, is used 
to swap the post-measurement states of qubits $n$ and $n+1$. The 
notation is identical to \fig{replaceinput}.}
    \label{fig:replaceoutput}
\end{figure}

The main idea behind this simulation is to use the circuit of \fig{replaceinput} in 
reverse, such as indicated 
in \fig{replaceoutput}. The main issue is that, in \fig{replaceinput}, we 
could use \eqs{id2}{id3} to swap the states of the qubits only because we knew upfront 
that one of the qubits being acted on was in the $\ket{0}$ state. Now, in 
\fig{replaceoutput}, we once again use these identities, but in the 
post-measurement states. 

The simulation is very similar to the one described previously for matchgate circuits 
with adaptive measurements. We begin by using the 
$\ket{+}$ ancilla to implement an arbitrary single-qubit gate on the final 
qubit, which effectively rotates the measurement basis, then we compute the 
probability of the corresponding measurement outcomes, Pr$(y_1 | \psi)$, using the 
method from the previous section. We sample classically according to the 
computed probabilities, fix the outcome of $y_1$, and replace the measurement in the 
circuit by the projector $P_{y_1}$. Since the state of that qubit after the 
measurement is either 
$\ket{0}$ or $\ket{1}$, we then use either $G(Z,X)$, as in \eq{id2}, or $G(-Z, 
X)$ [which satisfies an equation analogous to \eq{id2}, but when one of the 
inputs is in state $\ket{1}$] to swap the states of the final two qubits. 
We can now iterate this process to simulate the measurement of the last $k$ qubits, 
fixing the outcomes one by one, which consists of a weak simulation.

Although this procedure seems to only allow for the simulation of measurements on the last $k$ qubits, 
it is in fact completely general. Recall, from the discussion after \eq{JWtransf2}, that the overall ordering 
of the qubits is irrelevant. So, if the circuit we wish to simulate is not restricted to measurements of the 
last $k$ qubits, we can just map it into the fermionic picture, apply a permutation of the fermionic modes 
and map it back, resulting in an equivalent circuit in which the measured qubits are the last $k$ ones. 

This concludes the proof that matchgate circuits remain (weakly) simulable even after replacing measurements in the computational basis by arbitrary single-qubit measurements.

\section{Summary and open questions} \label{sec:summary}

We have shown that matchgates are classically simulable, in a strong sense, when 
the circuit acts on arbitrary product input states, includes an arbitrary number 
of intermediate measurements that condition the subsequent circuits, and is 
followed by measurement of an arbitrary subset of the output qubits, thereby 
generalizing previous known simulation results \cite{Valiant2002, 
Terhal2002,Jozsa2008b}. We have also shown how to include measurements of the 
qubits in rotated bases, but only by switching to a a weaker notion of simulation.

These results present an interesting parallel with other 
restricted models of computation. It is well known that complexity of simulation
cannot be attributed only to the allowed operations, but 
also to the allowed inputs and measurements, as well as the strength of the required 
simulation. Clifford circuits, for 
example, range from classically simulable, to universal for quantum computation, 
to $\#$P-hard to simulate (strongly) \cite{Jozsa2014, Koh2015}. Another example is 
linear optics, which can be classically simulated if the 
quasiprobability distribution of the input states and measurements satisfy 
certain conditions \cite{Bartlett2002, Veitch2013,Rahimi-Keshari2015}, is hard 
to simulate classically if Fock state inputs and number-resolving measurements 
are available \citep{Aaronson2013a}, and becomes universal for quantum computing 
if adaptive measurements are allowed \cite{Knill2001b}. In contrast to these 
examples, matchgates do not seem to gain any type of computational advantage 
from the addition of arbitrary product input states, even when adaptive 
measurements are allowed, and there is evidence that they do not gain any 
advantage from (single-qubit) non-computational-basis measurements either.

With these remarks in mind we pose a few open questions, both as continuations 
of the present work and as interesting investigations on the parallels between 
the different models:

\begin{itemize}[leftmargin=0.7 cm]
\item[(i)] Is it possible to extend the result of \sec{noncomputational} to 
allow for \emph{strong} simulation of measurements in arbitrary bases?
\item[(ii)] Although matchgates do not seem to benefit from arbitrary 
single-qubit inputs and measurements, we know that they become universal when 
certain multi-qubit input states or measurements are allowed. Is it possible to 
repeat the work done here, but to fully characterize the behaviour of matchgate 
circuits when supplemented with arbitrary two-qubit resources?
\item[(iii)] The matchgate simulation was extended to include periodic 
boundary conditions (i.e.\ extra matchgates between the first and last qubit) in 
the PI--SO setting in \cite{Brod2014}, and in the CI--MO setting in the \app. 
Can we also extend the result of \sec{main} to this geometry? Curiously, the 
circuit that is equivalent to \fig{replaceinput} for periodic boundary 
conditions corresponds to a geometry where matchgates are universal, as seen 
e.g.\ in Figure 4(b) of \cite{Brod2012}, although it might just use this 
geometry in a very restricted manner that does not break the simulability.
\item[(iv)] We have argued that, in the PI--SO setting, linear optics are 
classically simulable for the same reasons as matchgates, i.e.\ 
the linearity of \eq{linearoptics}. Can this parallel be extended further, to allow simulation of 
linear optics with inputs that are superpositions of photon numbers? What about 
to obtain bosonic versions of other matchgate results such as the mapping to 
logspace quantum computation \cite{Jozsa2010} or the compressed simulations of 
\cite{Kraus2011,Boyajian2013}?
\end{itemize}

\appendix*
\section{Fermionic transition amplitudes and Pfaffians} \label{sec:appendix}

In this Appendix, we give a few additional details on how determinants and 
Pfaffians arise in the fermionic transition amplitudes of \sec{backgCIMO}, 
following mostly along the steps of \cite{Terhal2002}. Throughout this Appendix we 
will be considering only a CI--MO scenario, where an input bit string $\ket{x}$ is 
acted on by some matchgate circuit $M$, and we wish to compute the corresponding 
outcome probabilities. We will restrict ourselves to the case where the circuit of 
matchgates preserves the number of ``fermions'', i.e.\ the Hamming weight of the bit 
strings. This corresponds to taking ${R'}=0$ in \eq{linearoptics}, but the argument 
is very similar in the more general case.

As an illustration, suppose initially we want to compute the transition amplitude 
between two 
$n$-bit strings $x$ and $y$ (i.e.\ all qubits are measured), given by $\bra{y}M\ket{x}$. Clearly this is non-zero only if $x$ and 
$y$ have the same Hamming weight, which we denote by $h$. Also, let indices 
$\{i_1, i_2, ... \, , i_h\}$ label the positions of the $h$ ones in $x$. Then, by 
recalling that $\creat{i}$ act as fermionic creation operators, we write

\begin{align*}
    \bra{y} M \ket{x} & = \bra{y} M \creat{i_1} \creat{i_2} ... \, \creat{i_h} \ket{\bar{0}_n} \\
    & = \sum_{p_1 ... \, p_h} \! \! \! \! R_{i_1, p_1} R_{i_2, p_2} ... \, R_{i_h, p_h}  \! \bra{y} \creat{p_1} \creat{p_2} ... \, \creat{p_h} 
    \ket{\bar{0}_n}.
\end{align*}
If similarly we use indices $\{ l_1, l_2,  ... \, , l_h \}$ to label the 
positions of the ones in $y$, it is easy to see that the only terms that survive 
in this sum are those for which the $p_j$'s are some permutation of the $l_j$'s. 
Furthermore, the anticommutation relations induce a minus sign on all odd 
permutations. This leads to the simple expression
\begin{equation}\label{eq:fermionsamplingAP}
    \bra{y} M \ket{x} = \det (R_{x,y}),
\end{equation}
where $R_{x,y}$ is an $h \times h$ submatrix of $R$ constructed as follows: 
first, make an $h \times n$ matrix $R_x$ by choosing the rows of $R$ that 
correspond to ones in $x$, and then construct $R_{x,y}$ by choosing the columns 
of $R_x$ that correspond to ones in $y$. Since the determinant of an $h \times 
h$ matrix, for $h\leq n$, can be computed in poly$(n)$ time, this gives a method 
for efficiently computing $\bra{y}M\ket{x}$.

Let us now consider the probabilities when only a subset of $k$ out of $n$ 
qubits is measured after the circuit $M$, which is what we actually need for the 
strong simulation of \thm{CIMO}. Note first that, for any given qubit $j$, we can 
write
\begin{subequations} \label{eq:projectors}
	\begin{align}
		\annih{j} \creat{j} & = \ketbra{0}{0}_j \\
		\creat{j} \annih{j} & = \ketbra{1}{1}_j.
	\end{align}
\end{subequations}
Remarkably, the measurement projectors themselves are quadratic in the fermionic 
operators, which has previously been identified as a crucial difference between 
quantum computing with fermionic and bosonic linear optics \cite{Knill2001a}. 
Let us proceed by again indexing the $h$ ones of $x$ by $\{i_1, i_2, ... \, , 
i_h\}$, and let $l_i$ indicate the position of the qubit assigned by the $i$th 
bit of $\tilde{y}$. We can then write
\begin{align} \label{eq:probystarAP}
\textrm{Pr}(\tilde{y}|x) & = \tr \bra{\tilde{y}}M\ketbra{x}{x}M^{\dagger}\ket{\tilde{y}} \notag \\
& = \bra{x} M^{\dagger} P_{\tilde{y}} M \ket{x}.
\end{align}
Here, $P_{\tilde{y}}$ is the projector $\ketbra{\tilde{y}}{\tilde{y}}$, which can be replaced
by a string of creation and annihilation operators where, for each index $l_i$, 
we chose $\annih{l_i} \creat{l_i}$ or 
$\creat{l_i} \annih{l_i}$ depending on whether $\tilde{y}_i$ is 0 or 1 [cf.\ the discussion
just after \eq{probystarPIMO}]. For example, for $\tilde{y} = 0 1 \ldots 0$ we would have
\begin{equation}
P_{\tilde{y}} = \left( \annih{l_1} \creat{l_1} \right) \left( \creat{l_2} \annih{l_2} \right)  ... \, \left( \annih{l_k} \creat{l_k}\right)
\end{equation}
Using \eq{linearoptics} (with ${R'}=0$) we can write
\begin{equation} \label{eq:preWick}
	\begin{aligned}
	\textrm{Pr}(\tilde{y}|x) = & \sum_{\textrm{$n$'s and $m$'s}} {R^{\textrm{*}}}_{l_1,m_1} R_{n_1,l_1} ... \, R_{l_k,m_k} R^{\textrm{*}}_{n_k,l_k} \\
	& \times \bra{\bar{0}_n} \annih{i_h} ... \, \annih{i_1} (\creat{n_1} \annih{m_1} ... \, \creat{m_k} \annih {n_k} ) \creat{i_1} ... \, \creat{i_h} \ket{\bar{0}_n}.
	\end{aligned}
\end{equation}

In order to simplify the above equation, usually one resorts to Wick's theorem, 
which provides a systematic way of rearranging creation and annihilation 
operators so as to reduce the expectation values in \eq{preWick} to complex numbers. We 
will not enter into the more arid details of Wick's theorem here, as the full 
procedure has already been carried out in \cite{Terhal2002}, we will just quote 
the final result:
\begin{equation}\label{eq:pfaffianAP}
\textrm{Pr}(\tilde{y}|x) = \textrm{Pf} (\tilde{M}).
\end{equation}
Here, $\tilde{M}$ is a $2(h+k) \times 2(h+k)$ antisymmetric matrix constructed 
in a specific manner from the matrix elements of $R$. The interested reader can 
find lookup tables with the rules for obtaining the matrix elements of 
$\tilde{M}$ in \cite{Terhal2002} (for a more direct relation between Wick's Theorem 
and Pfaffians, although in a somewhat different formalism, we direct the reader to \cite{Bravyi2005b}). 
The Pfaffian Pf$(A)$, that appears in 
\eq{pfaffianAP}, is a matrix polynomial related to the determinant. More 
specifically, if $A$ is an $N \times N$ antisymmetric matrix (as in our case), 
Pf$(A)$ is 0 if $N$ is odd, and for even $N$ it satisfies the relation
\begin{equation*}
\textrm{Pf} (A)^2 = \det(A).
\end{equation*}
Thus, once more the desired probabilities are given in terms of determinants of 
matrices constructed out of $R$, and thus can be computed efficiently. In fact, 
the original matchgate simulation of Valiant \cite{Valiant2002} exploited the 
fact that probabilities of matchgate circuits are given by Pfaffians, with no 
relation to fermions or Wick's theorem. Only later was this connection made 
explicit by Terhal and DiVincenzo in \cite{Terhal2002}.

One aspect of \thm{CIMO} introduced in \cite{Terhal2002} that was not found 
in \cite{Valiant2002} is simulation in the \emph{adaptive} setting. Let us give an 
idea why the simulation remains possible in this case. Suppose the circuit we wish to simulate consists of some 
initial matchgate circuit $M$ acting on an $n$-qubit state $\ket{x}$, followed by a 
measurement of the first qubit, $y_1$, and then one of two circuits $M_{1}$ or $M_{0}
$ corresponding to the two outcomes of $y_1$. We then 
wish to compute the probabilities of some $k$-qubit outcome $\tilde{y}_2$ on a subset 
of the $n-1$ remaining qubits. We can do this as follows:

\begin{itemize}[leftmargin=0.7 cm]
\item[(i)] Compute $\textrm{Pr}(y_1|x)$ using \eq{pfaffianAP} with $k=1$;
\item[(ii)] Classically sample according to the probabilities computed in (i), and fix $y_1$ accordingly.
\item[(iii)] We now wish to compute $\textrm{Pr}(\tilde{y}_2|x,y_1)$. To that end it suffices to compute
\begin{equation} \label{eq:probystarADAP}
\textrm{Pr}(\tilde{y}_2, y_1 | x) = \bra{x} M^{\dagger} P_{y_1} M_{y_1}^{\dagger} P_{\tilde{y}_2} M_{y_1} P_{y_1} M \ket{x},
\end{equation}
where $P_{y_1}$ is $\annih{1} \creat{1}$ or $\creat{1} \annih{1}$ if $y_1$ is 0 
or 1 respectively. It is clear that this expression is amenable to exactly the 
same treatment in terms of Wick's theorem as \eq{probystarAP}. In 
\cite{Terhal2002} it is also shown how to rewrite this expression as the 
Pfaffian of some efficiently-computable antisymmetric square matrix.
\end{itemize}

Clearly, steps (i)-(iii) can be generalized to allow for any number of rounds of 
intermediate measurements, with any number of qubits being measured in each 
round.

One small extension of these arguments follows directly from the work of 
\cite{Terhal2002}, although it does not seem to be pointed out anywhere: efficient 
classical simulation remains possible even if we allow for 
``periodic boundary conditions'', that is, if we also allow matchgates to act 
between the first and last qubits. To see that, note that we can write, for 
example,
\begin{equation*}
X_1 X_n = - \left ( \prod_{i=1}^{n} Z_i \right ) Y_1 Z_2 Z_3 \ldots Z_{n-1} Y_n,
\end{equation*}
with equivalent equations for the other matchgate generators of 
\subeq{JWtransf2}. But $\prod_{i=1}^{n} Z_i$ is just the operator that measures 
the overall parity of the whole $n$-qubit state. Since a circuit of matchgates preserves the parity of the initial state, whenever the input is in the computational basis this operator can just be replaced by $+1$ or $-1$ depending on the parity of the input. Also recall that any gate 
generated by a Hamiltonian of the type $Y_1 Z_2 Z_3 \ldots Z_{n-1} Y_n$ can be 
decomposed into a circuit of O$(n^2)$ nearest-neighbour matchgates \cite{Jozsa2008b}. Thus, any  matchgate circuit with periodic boundary conditions can be replaced by a
circuit of nearest-neighbour matchgates that has the same action on that input 
state, with only polynomial overhead, and thus \thm{CIMO} still holds. 
Remarkably, this is the only type of non-nearest-neighbour matchgate we can add 
to the set without leading to universal quantum computation \cite{Brod2014}.

\begin{acknowledgments}
The author would like to thank R.~Jozsa and D.~Gottesman for valuable feedback and helpful discussions, and J.~Emerson for suggesting the extension of initial results to also include arbitrary single-qubit measurements. This research was supported by the Perimeter Institute for Theoretical Physics. Research at Perimeter Institute is supported by the Government of Canada through Industry Canada and by the Province of Ontario through the Ministry of Research and Innovation.
\end{acknowledgments}

\bibliography{matchgates2}
	
\end{document}